\documentclass[11pt]{article}

\usepackage{fullpage}

\usepackage{algorithm}
\usepackage{algpseudocode}

\usepackage{booktabs} 
\usepackage{lscape}   

\usepackage{cite}

\usepackage{latexsym}
\usepackage{url}
\usepackage{listings}
\usepackage{xspace}
\usepackage{color}
\usepackage{rotating,graphicx}

\usepackage{amsmath}
\usepackage{amssymb}
\usepackage[binary-units]{siunitx}

\newcommand{\eg}{e.g.\@\xspace}

\newcommand{\ceiling}[1]{\lceil #1\rceil}
\newcommand{\floor}[1]{\lfloor #1\rfloor}

\newcommand{\skips}[1]{\mathtt{skips[}#1\mathtt{]}}

\newcommand{\mpibcast}{\textsf{MPI\_Bcast}\xspace}
\newcommand{\mpiallgatherv}{\textsf{MPI\_Allgatherv}\xspace}

\newcommand{\bidirec}[2]{\textsf{Send}(#1)\parallel\textsf{Recv}(#2)\xspace}

\newtheorem{theorem}{Theorem}
\newtheorem{lemma}{Lemma}

\newcommand{\repetitions}{35}

\newcommand{\gcc}{\texttt{gcc~8.3.0}\xspace}
\newcommand{\hydrampich}{\texttt{mpich}~3.3\xspace}
\newcommand{\hydraopenmpi}{OpenMPI~4.0.5\xspace}
\newcommand{\hydraintelmpi}{IntelMPI~2019.9\xspace}
    
\DeclareSIUnit{\Gbps}{\giga\bit/\s}
\DeclareSIUnit{\microsecond}{\SIUnitSymbolMicro s}

\title{(Poly)Logarithmic Time Construction of Round-optimal $n$-Block
  Broadcast Schedules for Broadcast and irregular Allgather in MPI\thanks{These results were announced briefly at SPAA 2022~\cite{Traff22:bcastba} and in fuller form at CLUSTER 2022~\cite{Traff22:bcast}}}
\author{Jesper Larsson Tr\"aff\\
  TU Wien\\
  Faculty of Informatics\\
  Institute of Computer Engineering, Research Group Parallel Computing 191-4\\
  Treitlstrasse 3, 5th Floor, 1040 Vienna, Austria}

\begin{document}
\maketitle

\begin{abstract}
  We give a fast(er), communication-free, parallel construction of
  optimal communication schedules that allow broadcasting of $n$
  distinct blocks of data from a root processor to all other
  processors in $1$-ported, $p$-processor networks with fully
  bidirectional communication. For any $p$ and $n$, broadcasting in
  this model requires $n-1+\ceiling{\log_2 p}$ communication rounds.
  In contrast to other constructions, all processors follow the same,
  circulant graph communication pattern, which makes it possible to
  use the schedules for the allgather (all-to-all-broadcast) operation
  as well. The new construction takes $O(\log^3 p)$ time steps per
  processor, each of which can compute its part of the schedule
  independently of the other processors in $O(\log p)$ space. The
  result is a significant improvement over the sequential $O(p \log^2
  p)$ time and $O(p\log p)$ space construction of Tr\"aff and Ripke
  (2009) with considerable practical import.  The round-optimal
  schedule construction is then used to implement communication
  optimal algorithms for the broadcast and (irregular) allgather
  collective operations as found in MPI (the \emph{Message-Passing
    Interface}), and significantly and practically improves over the
  implementations in standard MPI libraries (\texttt{mpich}, OpenMPI,
  Intel MPI) for certain problem ranges.  The application to the
  irregular allgather operation is entirely new.
\end{abstract}

\section{Introduction}

The problem is the following. In a network of $p$ processors, one
designated \emph{root processor} has $n$ blocks of data that have to be
broadcast (transmitted to) all other processors. The processors are
fully connected, and in one communication operation, a processor can
simultaneously receive one block of data from one other processor and
send a block of data to one other (possibly different) processor. This
is the $1$-ported, fully connected, bidirectional send-receive
model~\cite{BarNoyKipnisSchieber00}.  The lower bound for the
broadcast operation under these assumptions is $n-1+\ceiling{\log_2
  p}$ \emph{communication rounds} as is well known, see
\eg~\cite{BarNoyHo99}. An optimal \emph{broadcast schedule} reaches
this lower bound, and for each processor explicitly specifies for each
round which of the $n$ blocks are to be sent and received to and from
which other processors. In a homogeneous, linear-cost communication
model, where transferring a message of $m$ units between any two
processors takes $\alpha+\beta m$ units of time, this lower bound
gives rise to a broadcast time of $\alpha\ceiling{\log_2 p-1} +
2\sqrt{\ceiling{\log_2 p-1}\alpha\beta m}+ \beta m$ by dividing $m$
appropriately into $n$ blocks.

It is well-known that the problem can be solved optimally when $p$ is
a power of $2$, and many hypercube and binary butterfly algorithms
exist, see for instance the pioneering work on EDBT (Edge Disjoint
Binomial Trees) of~\cite{JohnssonHo89}. These algorithms are typically
difficult to generalize to arbitrary numbers of processors; an
exception to this claim is the appealing construction based on a
hypercube algorithm given in~\cite{Jia09}. The preprocessing required
for the schedule construction is $O(\log p)$ time steps per processor,
but different processors have different roles and different
communication patterns in the construction.  A different, explicit,
optimal construction with a symmetric communication pattern was given
in~\cite{Traff08:optibcast}, however with a prohibitive, sequential
preprocessing time of $O(p\log^2 p)$. Both of these
algorithms~\cite{Jia09,Traff08:optibcast} have been explicitly used in
MPI (the \emph{Message-Passing Interface}, the \emph{de facto}
standard for programming applications for high-performance,
distributed memory systems~\cite{MPI-3.1}) libraries for implementing
the \mpibcast operation~\cite{Traff06:mpisxcoll}. Other optimal, less
explicit constructions where given
in~\cite{BarNoyKipnisSchieber00,KwonChwa95}, see also
\cite{GregorSkrekovskiVukasinovic18}. In the different $LogP$
communication cost model, also assuming a fully-connected
communication network, an optimal result for $k$-item broadcast was
claimed in~\cite{KarpSahaySantosSchauser93,Santos99}. In arbitrary,
non-fully connected communication structures, the problem of finding
minimum round broadcast communication schedules is
NP-complete~\cite{Farley80,HarutyunyanShao06,HarutyunyanLi20}.

The construction presented in the present paper improves on the sequential
$O(p \log^2 p)$ algorithm in~\cite{Traff08:optibcast} by giving a
construction that for any processor in $O(\log^3 p)$ time steps
computes its necessary send- and receive schedules. The construction
uses the same, symmetric communication pattern
as~\cite{Traff08:optibcast}, which makes it possible to use the
schedules also for implementing a all-to-all-broadcast operation (also
known as allgather) in which each processor has data of possibly
different sizes to broadcast to all other processors. In MPI, this
operation is called \mpiallgatherv. It is not obvious whether this is
possible with other schedule constructions in which the processors
have different roles and communication patterns~\cite{Jia09}.


\section{The Schedule Computation}

In the following, $p$ will denote the number of processors which will
be consecutively ranked $0,1,\ldots, p-1$, such that each of the $p$
processors knows its own \emph{rank} $r\in\{0,1,\ldots,p-1\}$. Let
furthermore $q=\ceiling{\log_2 p}$. The communication structure for
the broadcast and all-to-all-broadcast algorithms will be a
\emph{circulant graph} $G=(V,E)$ with vertices $V=\{0,\ldots,p-1\}$
and (directed) to- and from-edges $(r,t^k_r),(f^k_r,r)\in E$ with
$t^k_r = (r+\skips{k})\bmod p$ and $f^k_r = (r-\skips{k}+p)\bmod p$
for each processor $r\in V$. The skips (or jumps) in the circulant
graph, $\skips{k}, 0\leq k<q$, are computed by successive halving of
$p$ as shown in Algorithm~\ref{alg:circulants}. When $p=2^q$ is a
power of two, the skips are likewise powers of two,
$1,2,4,\ldots,2^{q-1}$. For convenience, we compute also
$\skips{q}=p$. With these roughly doubling skips (jumps), the diameter
of $G$ is clearly $q$.

\begin{algorithm}
  \caption{Computing the skips (jumps) for $p$ processor circulant
    graph ($q=\ceiling{\log_2 p}$).}
  \label{alg:circulants}
\begin{algorithmic}[1]
  \State $k\gets q$
  \While{$p>1$}
  \State $\skips{k}\gets p$
  \State $p\gets \floor{p/2}+p\bmod 2$ \Comment which is $\ceiling{p/2}$
  \State $k\gets k-1$
  \EndWhile
  \State $\skips{k}\gets p$  
\end{algorithmic}
\end{algorithm}

From the construction of the skips (jumps) in
Algorithm~\ref{alg:circulants}, the following properties can easily be
observed.

\begin{lemma}
  \label{lem:sizes}
  For any $k, 0\leq k<q$ it holds that $\skips{k} + \skips{k} \geq
  \skips{k+1}$ (equivalently that
  $\skips{k}\geq\skips{k+1}-\skips{k}$) and (by induction, observing
  that $\skips{0}=1$ and $\skips{1}=2$) that $\sum_{i=0}^k\skips{i}
  \geq \skips{k+1}-1$. It trivially (since $\skips{q}=p$ and
  $\skips{0}=1$) holds that $\sum_{k=0}^{q-1}(\skips{k+1}-\skips{k}) =
  p-1$.
\end{lemma}

The broadcast (and allgather) algorithm(s) will divide their data into
$n$ roughly equal sized blocks, and go through $n-1+\ceiling{\log_ 2
  p}=n-1+q$ communication rounds. In round $i, x\leq i<x+n-1+q$ for
some round offset $x$ (to be explained later), each processor will
send and receive data blocks on its to- and from-edges
$(r,r+\skips{i\bmod q})\bmod p$ and $((r-\skips{i\bmod q}+p)\bmod
p,r)$. For the algorithm to be correct and optimal, each processor must
in each round a) send a block that it already had (received in a
previous round), and b) receive a new block not seen so far.

\begin{algorithm}
  \caption{Finding the baseblock for processor $r,0<r<p$.}
  \label{alg:baseblock}
\begin{algorithmic}[1]
  \Function{baseblock}{$r$}
  \State $k\gets q$
  \While{$r\neq \skips{k}$}
  \State $k\gets k-1$
  \If{$\skips{k}<r$}
  \State $r\gets r-\skips{k}$
  \EndIf
  \EndWhile
  \State \Return{$k$}
  \EndFunction
\end{algorithmic}
\end{algorithm}

Without loss of generality, it will be assumed that broadcasting is
done from \emph{root processor} $r=0$ (otherwise, virtually renumber
the processors by subtracting the rank of the root $r\neq 0$ from the
processor numbers modulo $p$). The root processor has the $n$ blocks
of data initially, and sends these successively $0,1,2,\ldots,n-1$ to
processors $(r,r+\skips{i\bmod q})\bmod p$. When a processor $r'\neq
r$ receives a first block in some round $i$, it greedily sends this block
further along the edges $(r',r'+\skips{i'\bmod q})\bmod p$ in the
following rounds $i'=i+1,i+2,\ldots,q-1$. This is the best that can be
done in the first $q$ rounds where no processors except the root has
any blocks of data.  The first block that a processor receives is
called its \emph{baseblock}. The list of baseblocks for processors
$r=1,2,\ldots,p-1$ can be computed in linear time by observing what
the recipe does: In round $i$, root processor $r=0$ sends a new block
to processor $\skips{i}$, and all processors $1\leq r<\skips{i}$ send
their baseblocks to processors $r+\skips{i}$ (when this is smaller
than $\skips{i+1}$). This gives rise to patterns as shown for $p=20$
in Table~\ref{tab:p20}, and for $p=33,32,31$ in Tables~\ref{tab:p33},
\ref{tab:p32} and~\ref{tab:p31}. For processor $r$ to determine its
baseblock, a linear time procedure is much too slow;
Algorithm~\ref{alg:baseblock} shows how to compute the baseblock for
each non-root processor $r, r>0$ more efficiently by searching through
the skips (jumps) starting from $\skips{q}$.  Obviously, finding the
baseblock for processor $r$ takes $O(q)$ steps. To see that the
algorithm is also correct, first observe that the baseblock for a
processor $r$ with $r=\skips{k}$ is $k$. The algorithm therefore
searches for the largest $k$ such that $\skips{k}\leq r$. This is the
value of $k$ such that processor $r$ lies in the range
$[\skips{k},\skips{k+1}-1]$; we call this range the \emph{homerange}
for $r$. Since the list of baseblocks in the homerange excluding
processor $\skips{k}$ is the same as the list of baseblocks in the
range $[1,\skips{k}-1]$, the baseblock for processor $r$ can be found
by searching in this range, which the algorithm accomplishes by
subtracting $\skips{k}$ from $r$. 

\begin{algorithm}
  \caption{Finding all blocks in a range $[a,b]$, assuming $0\leq
    a\leq b<p$ The function will be called also for cyclic ranges $[a,p-1]$
    followed by $[0,b]$, and can easily be modified for this
    situation.}
  \label{alg:rangeblocks}
\begin{algorithmic}[1]
  \Function{RangeBlocks}{$a,b,k$}
  \While{$b<\skips{k}$}
  \State $k\gets k-1$
  \EndWhile
  \State $k\gets k+1$
  \State $k'\gets k$
  \While{$a\leq\skips{k'}$}
  \State $k'\gets k'-1$
  \EndWhile
  \State $k'\gets k'+1$
  \Comment It holds that $a\leq\skips{k'}$ and $b<\skips{k}$ with
  smallest $k,k'$
  \If{$k'=k$}
  \State $B\gets \Call{RangeBlocks}{a-\skips{k-1},b-\skips{k-1},k}$
  \ElsIf{$k'+1=k$}
  \State $B\gets\{k'\}$
  \If{$a<\skips{k'}$}
  \State
  $B\gets B\cup \Call{RangeBlocks}{a-\skips{k'-1},\skips{k'}-\skips{k'-1}-1,k}$
  \EndIf
  \State $B\gets B\cup\Call{RangeBlocks}{0,b-\skips{k'},k'}$
  \Else \Comment Possible exceptions for $k=1,2,3$ need to be handled explicitly
  \State $B\gets \{0,1,\ldots,k-1\}$
  \EndIf
  \State \Return{$B$}
  \EndFunction
\end{algorithmic}
\end{algorithm}

For the full broadcast operation, the $n-1+q$ rounds are divided into
phases of $q$ rounds. To ensure that this can always be done, an
offset $x$ is simply computed such that $x+n-1+q$ is a multiple of
$q$. In the first phase, the baseblocks are distributed from the root
processor as explained above. In the next phase, each processor $r$
will in round $i$ receive blocks from processors $(r-\skips{i\bmod
  q}+p)\bmod p$ that were received in the previous phase. Each block
thus received, must be a block that was indeed received by the
from-processor. For processor $r$ to compute which blocks could have
been received by a from-processor, a function is needed that can
compute for any successive range of processors $[a,b]$ which
baseblocks are in that range.  This is accomplished by the recursive
Algorithm~\ref{alg:rangeblocks} in $O(q)$ steps. The algorithm uses
the skips (jumps) to guide the search for new baseblocks based on the
observation that in a homerange $[\skips{k},\skips{k+1}-1]$ all $k+1$
blocks $k,k-1,\ldots,0$ are present (with the possible exception of
the ranges $[\skips{1},\skips{2}-1]$, $[\skips{2},\skips{3}-1]$
depending on the magnitude of the skips. For instance, in the example
in Table~\ref{tab:p33} for $p=33$, the range
$[3,4]=[\skips{2},\skips{3}-1]$ has only baseblocks $2,0$). Since the
algorithm goes through the skips once, the $O(q)$ time bound follows;
here, one needs to observe that the call to \textsc{rangeblocks}
starting from $a=0$ does not lead to a doubling of the number of
recursive calls. Starting from $k=q$ the algorithm locates the
smallest $k$ and $k'$ with $k'\leq k$ such that $\skips{k'}\geq a$ and
$\skips{k}>b$. There are here two possibilities to consider. Either
$[a,b]<\skips{k}$ in which case the baseblocks can be found by
searching the range $[a-\skips{k},b-\skips{k}]$, or
$a\leq\skips{k'}\leq b<\skips{k}$. In the latter case, either no
complete homerange is contained in $[a,b]$ in which case the search is
done on the two subranges $[a,\skips{k'}-1]$ and $[0,b-\skips{k'}]$,
or at least the homerange $[\skips{k'},\skips{k'+1}-1]$ (precisely,
the range $[\skips{k'},\skips{k-1}-1]$) is contained in $[a,b]$ and
all baseblocks $0,1,\ldots k-1$ are in the range, and no further
searches are needed since no larger blocks than $k-1$ can exist in
$[a,b]$.

The \textsc{rangeblocks} computation is for each processor now used to
compute arrays \texttt{sendsched} and \texttt{recvsched}, such that in
round $i, 0\leq i<q$ of each phase, the processor sends a block
$\mathtt{sendsched}[i]$ and receives a block
$\mathtt{recvsched}[i]$. Obviously, it must hold that
$\mathtt{sendsched}[i]=\mathtt{recvsched}[i']$ for some $i',0\leq
i'<i$, that is, the block that a processor is about to send in round
$i$ has been received in some previous round $i'<i$. These two arrays
comprise the send and receive schedules for processor $r$.
Algorithm~\ref{alg:receiveschedule} computes first the
\texttt{recvsched}.

\begin{algorithm}
  \caption{Computing the first $k$ blocks of the receive schedule
    \texttt{recvsched} for processor $r$.}
  \label{alg:receiveschedule}
  \begin{algorithmic}[1]
    \Function{recvsched}{$r,k$}
    \Comment Return the first $k, k\leq q$ blocks for processor $r$
    \State $B\gets\emptyset$ \Comment No blocks so far
    \For{$i=0,1,\ldots,k$}
    \If{$\skips{i}\leq r<\skips{i+1}$}
    \State$\mathtt{recvsched}[i]\gets \Call{baseblock}{r}$
    \Else
    \If{$i=0$}
    \State $b\gets\Call{baseblock}{(r-1+p)\bmod p}$
    \ElsIf{$i<q-1$}
    \Comment Check for new block in range of size $\skips{i+1}-\skips{i}$
    \State $U\gets\Call{rangeblocks}{r-\skips{i+1}+1,r-\skips{i},q}$
    \If{$U\setminus B=\emptyset$}
    \State $U\gets\Call{rangeblocks}{r-\sum_{j=0}^i \skips{j},r-\skips{i+1},q}$
    \EndIf
    \State $b\gets \max(U\setminus B)$
    \Else \Comment Last block $i=q-1$
    \State $b\gets\{0,\ldots,q-1\}\setminus B$
    \EndIf
    \State $B\gets B\cup\{b\}$
    \State $\mathtt{recvsched}[i] \gets b-q$
    \EndIf
    \EndFor
    \State \Return \texttt{recvsched}
    \Comment Receive blocks for first $k$ rounds
    \EndFunction
\end{algorithmic}
\end{algorithm}

Processor $r, r>0$ will in round $i$ receive a block from the previous
phase from processor $r'=(r-\skips{i\bmod q}+p)\bmod p$. This must be
a new block that processor $r$ has not already received in a previous
round (by a processor closer to $r$), and a block that processor $r'$
already has. The algorithm maintains the set of blocks $B$ that
processor $r$ already has received in the phase. A new block from
processor $r'$ can be in the range of baseblocks
$[r-\skips{i+1}+1,r']$ (omitting now for readability the modulo $q$
and $p$ computations), since processor $r$ will in the next round
$i+1$ receive a block from processor $r-\skips{i+1}$. This range
is searched with the \textsc{rangeblock} algorithm for a set of blocks
$U$. If this does not give rise to a new block for processor $r$, the
range over processors from which $r'$ can possibly have received
blocks in the preceding $i-1$ rounds (see Lemma~\ref{lem:sizes}) is
searched. This range is
$[r-\sum_{j=0}^{i}\skips{j},r-\skips{i+1}]$. If more than one possible
new block is found, $U\setminus B\neq\emptyset$, the largest block is
chosen. The construction for each of the $k$ rounds invokes the
\textsc{rangeblock} computation for a worst-case complexity of $O(kq)$
time steps.

The correctness of the construction, guaranteeing that a new, unique
block is received in every round in a phase is based on the following
two observations.
\begin{lemma}
  \label{lem:ranges}
  In any range of processors $[a,b]$ of size $\skips{k}$
  there are at least $k$ different baseblocks.
  In any range of processors $[a,b]$ of size $1+\sum_{i=0}^{k-1}\skips{i}$
  there are at least $k+1$ different baseblocks.
\end{lemma}
We argue for this key lemma by considering ranges of size
$\skips{k+1}-\skips{k}$ which should have at least $k$ different
baseblocks. A range of size $1$ (as is the case for $k=0$) has one
block; also any two successive baseblocks are always different. For
$k\geq 3$, a range of size $\skips{k+1}-\skips{k}$ starting at a
processor rank $i=\skips{k'}$ for $k'\geq k$ has $k+1$ different
baseblocks for $k\geq 3$, since the baseblock at rank $i$ is $k'$,
and the following blocks in the range are the same as the baseblocks
for the processors in the range $1,\ldots \skips{k}-1$, which by
construction has the baseblocks from $0$ to $k-1$. If we shift the
range ``upwards'' (for a new range starting now at rank $i+1$) the
first baseblock $k'$ will disappear. In the case where
$\skips{k}+\skips{k}=\skips{k+1}$ the ``new'' baseblock at the end of
the range will be $k+1$ which is not in the range, and therefore there
will again be in total $k+1$ different baseblocks. Otherwise, there
will one baseblock less in the range, that is $k$ different blocks in
total. Shifting further up, this will not change, until arriving at a
rank $i'+1$ which again corresponds the baseblocks in range
$1,\ldots\skips{k}-1$, where there will again be $k+1$ different baseblocks.
The full proof requires a careful case analysis.

The schedule in Table~\ref{tab:p33} in particular illustrates the
claims of the lemma; for instance by sliding the range of size
$4=9-5=\skips{4}-\skips{3}$ upwards. Up to starting rank $9$ the range
has $4$ different baseblocks, whereas starting from rank $10$ the
range has $3$ baseblocks. It suffices to consider ``upwards'' shifts
of the ranges by virtually doubling the number of processors.  The
schedule for $p=2^5=32$ in Table~\ref{tab:p32} on the other hand shows
a particular property for powers of two, namely that the number of
different blocks in a range of size $2^k=\skips{k+1}-\skips{k}$ is
exactly $k+1$. This property, among others, makes it relatively easy
to compute the schedules for $p=2^q$ processors in $O(q)$ time steps,
as is well-known.

\begin{algorithm}
  \caption{Computing the send schedule \texttt{sendsched} for
    processor $r$ (straightforward variant).}
  \label{alg:sendschedule}
\begin{algorithmic}[1]
  \Function{sendsched}{$r$}
  \For{$i=0,\ldots,q-1$}
  \State $\texttt{sendsched}[i]\gets\Call{recvsched}{r+\skips{i},i}[i]$
  \EndFor
  \State\Return\texttt{sendsched}
  \EndFunction
\end{algorithmic}
\end{algorithm}

The send schedule computation for processor $r$ in a straightforward,
but obviously non-optimal way reuses the receive schedule computation
and is shown as Algorithm~\ref{alg:sendschedule}. For each round
$i,0\leq i<q$, it invokes the receive block computation for processor
$r+\skips{i}$ and takes the $i$th receive block as the send
block. This obviously fulfills the correctness condition. The
complexity is bounded by the complexity of the receive block
computation for $i, i=0,\ldots, q-1$ blocks which amounts to
$O(\sum_{i=0}^{q-1} iq)=O(\log^3 p)$.

Summarizing the discussion, we claim the following theorem.

\begin{theorem}
  In a fully connected, $1$-ported communication network with $p$
  processors, $q=\ceiling{\log_2 p}$, a correct communication schedule
  for round-optimal broadcasting in a symmetric, circulant graph
  pattern that for each processor $r,0\leq r<p$,
  consists of two arrays of size $q$ of send and receive blocks can
  be computed in $O(\log^3 p)$ time steps.
\end{theorem}

By the way the \texttt{recvsched} array for processor $r$ is constructed
by Algorithm~\ref{alg:receiveschedule}, it has the property of
containing the baseblock for $r$ (once), and $q-1$ different blocks
from the previous phase (thus negative). Thus, by using the schedule
over a sequence of phases, in the last phase each processor will have
received all blocks from previous phases, and one new baseblock for
the current phase. When the number of rounds and phases is finite, the
last phase can cap all blocks beyond $n$ to the last block $n-1$. This
will guarantee that after the last phase, all blocks have been
received by all processors. In Section~\ref{sec:bcastalg} this
observation shows the correctness and optimality of the $n$ block
broadcast algorithm.

\subsection{Examples}

\begin{table*}
\begin{center}
  \caption{Baseblocks, receive and send schedules for $p=20$ for all
    processors $r,0\leq r<p$. The jumps in the circulant graph are
    $\skips{i}=1,2,3,5,10,20$ indicated by the vertical lines.}
  \label{tab:p20}
  \begin{small}
  \begin{tabular}{lc|c|c|cc|ccccc|cccccccccc}
    \toprule
Procss $r$ &  0 & 1 & 2 & 3 & 4 & 5 & 6 & 7 & 8 & 9 & 10 & 11 & 12 & 13 & 14 & 15 & 16 & 17 & 18 & 19 \\
Baseblock & 0 & 0 & 1 & 2 & 0 & 3 & 0 & 1 & 2 & 0 & 4 & 0 & 1 & 2 & 0 & 3 & 0 & 1 & 2 & 0 \\
\midrule
& -5 & 0 & -5 & -4 & -3 & -5 & -2 & -5 & -4 & -3 & -5 & -1 & -5 & -4 & -3 & -5 & -2 & -5 & -4 & -3 \\
& -3 & -3 & 1 & -5 & -4 & -3 & -3 & -2 & -5 & -4 & -3 & -3 & -1 & -5 & -4 & -3 & -3 & -2 & -5 & -4 \\
Recv schedule & -4 & -4 & -3 & 2 & 0 & -4 & -4 & -3 & -2 & -2 & -4 & -4 & -3 & -1 & -1 & -4 & -4 & -3 & -2 & -2 \\
& -2 & -2 & -2 & -2 & -2 & 3 & 0 & 1 & 2 & 0 & -2 & -2 & -2 & -2 & -2 & -1 & -1 & -1 & -1 & -1 \\
& -1 & -1 & -1 & -1 & -1 & -1 & -1 & -1 & -1 & -1 & 4 & 0 & 1 & 2 & 0 & 3 & 0 & 1 & 2 & 0 \\ 
\midrule
& 0 & -5 & -4 & -3 & -5 & -2 & -5 & -4 & -3 & -5 & -1 & -5 & -4 & -3 & -5 & -2 & -5 & -4 & -3 & -5 \\
& 1 & -5 & -4 & -3 & -3 & -2 & -5 & -4 & -3 & -3 & -1 & -5 & -4 & -3 & -3 & -2 & -5 & -4 & -3 & -3 \\ 
Send schedule & 2 & 0 & -4 & -4 & -3 & -2 & -2 & -4 & -4 & -3 & -1 & -1 & -4 & -4 & -3 & -2 & -2 & -4 & -4 & -3 \\
& 3 & 0 & 1 & 2 & 0 & -2 & -2 & -2 & -2 & -2 & -1 & -1 & -1 & -1 & -1 & -2 & -2 & -2 & -2 & -2 \\
& 4 & 0 & 1 & 2 & 0 & 3 & 0 & 1 & 2 & 0 & -1 & -1 & -1 & -1 & -1 & -1 & -1 & -1 & -1 & -1 \\
\bottomrule
  \end{tabular}
  \end{small}
  \end{center}
\end{table*}

\noindent
\begin{sidewaystable*}
  \begin{center}
    \caption{Receive and send schedules for $p=33$
      ($\skips{i}=1,2,3,5,9,17,33$, vertical lines) for all processors
      $r,0\leq r<p$.}
\label{tab:p33}
    \begin{scriptsize}
      \begin{tabular}{c|c|c|cc|cccc|cccccccc|cccccccccccccccccc}
        \toprule
0 & 1 & 2 & 3 & 4 & 5 & 6 & 7 & 8 & 9 & 10 & 11 & 12 & 13 & 14 & 15 & 16 & 17 & 18 & 19 & 20 & 21 & 22 & 23 & 24 & 25 & 26 & 27 & 28 & 29 & 30 & 31 & 32 \\
-1 & 0 & 1 & 2 & 0 & 3 & 0 & 1 & 2 & 4 & 0 & 1 & 2 & 0 & 3 & 0 & 1 & 5 & 0 & 1 & 2 & 0 & 3 & 0 & 1 & 2 & 4 & 0 & 1 & 2 & 0 & 3 & 0 \\
\midrule
-6 & 0 & -6 & -5 & -4 & -6 & -3 & -6 & -5 & -4 & -2 & -6 & -5 & -4 & -6 & -3 & -6 & -5 & -1 & -6 & -5 & -4 & -6 & -3 & -6 & -5 & -4 & -2 & -6 & -5 & -4 & -6 & -3 \\
-3 & -3 & 1 & -6 & -5 & -4 & -4 & -3 & -6 & -5 & -4 & -2 & -6 & -5 & -4 & -4 & -3 & -6 & -5 & -1 & -6 & -5 & -4 & -4 & -3 & -6 & -5 & -4 & -2 & -6 & -5 & -4 & -4 \\
-4 & -4 & -3 & 2 & 0 & -5 & -5 & -4 & -3 & -3 & -5 & -4 & -2 & -2 & -5 & -5 & -4 & -3 & -3 & -3 & -1 & -1 & -5 & -5 & -4 & -3 & -3 & -5 & -4 & -2 & -2 & -5 & -5 \\
-2 & -2 & -4 & -3 & -3 & 3 & 0 & 1 & 2 & -6 & -3 & -3 & -3 & -3 & -2 & -2 & -2 & -2 & -4 & -4 & -3 & -3 & -1 & -1 & -1 & -1 & -6 & -3 & -3 & -3 & -3 & -2 & -2 \\
-1 & -5 & -2 & -2 & -2 & -2 & -2 & -2 & -2 & 4 & 0 & 1 & 2 & 0 & 3 & 0 & 1 & -4 & -2 & -2 & -2 & -2 & -2 & -2 & -2 & -2 & -1 & -1 & -1 & -1 & -1 & -1 & -1 \\
-5 & -1 & -1 & -1 & -1 & -1 & -1 & -1 & -1 & -1 & -1 & -1 & -1 & -1 & -1 & -1 & -1 & 5 & 0 & 1 & 2 & 0 & 3 & 0 & 1 & 2 & 4 & 0 & 1 & 2 & 0 & 3 & 0 \\
\midrule
0 & -6 & -5 & -4 & -6 & -3 & -6 & -5 & -4 & -2 & -6 & -5 & -4 & -6 & -3 & -6 & -5 & -1 & -6 & -5 & -4 & -6 & -3 & -6 & -5 & -4 & -2 & -6 & -5 & -4 & -6 & -3 & -6 \\
1 & -6 & -5 & -4 & -4 & -3 & -6 & -5 & -4 & -2 & -6 & -5 & -4 & -4 & -3 & -6 & -5 & -1 & -6 & -5 & -4 & -4 & -3 & -6 & -5 & -4 & -2 & -6 & -5 & -4 & -4 & -3 & -3 \\
2 & 0 & -5 & -5 & -4 & -3 & -3 & -5 & -4 & -2 & -2 & -5 & -5 & -4 & -3 & -3 & -3 & -1 & -1 & -5 & -5 & -4 & -3 & -3 & -5 & -4 & -2 & -2 & -5 & -5 & -4 & -4 & -3 \\
3 & 0 & 1 & 2 & -6 & -3 & -3 & -3 & -3 & -2 & -2 & -2 & -2 & -4 & -4 & -3 & -3 & -1 & -1 & -1 & -1 & -6 & -3 & -3 & -3 & -3 & -2 & -2 & -2 & -2 & -4 & -3 & -3 \\
4 & 0 & 1 & 2 & 0 & 3 & 0 & 1 & -4 & -2 & -2 & -2 & -2 & -2 & -2 & -2 & -2 & -1 & -1 & -1 & -1 & -1 & -1 & -1 & -1 & -5 & -2 & -2 & -2 & -2 & -2 & -2 & -2 \\
5 & 0 & 1 & 2 & 0 & 3 & 0 & 1 & 2 & 4 & 0 & 1 & 2 & 0 & 3 & 0 & -5 & -1 & -1 & -1 & -1 & -1 & -1 & -1 & -1 & -1 & -1 & -1 & -1 & -1 & -1 & -1 & -1 \\
    \bottomrule
      \end{tabular}
\end{scriptsize}
  \end{center}

\begin{center}
\caption{Receive and send schedules for $p=32$
  ($\skips{i}=1,2,4,8,16,32$, vertical lines) for all processors
  $r,0\leq r<p$.}
\label{tab:p32}
  \begin{scriptsize}
  \begin{tabular}{c|c|cc|cccc|cccccccc|cccccccccccccccccc}
    \toprule
0 & 1 & 2 & 3 & 4 & 5 & 6 & 7 & 8 & 9 & 10 & 11 & 12 & 13 & 14 & 15 & 16 & 17 & 18 & 19 & 20 & 21 & 22 & 23 & 24 & 25 & 26 & 27 & 28 & 29 & 30 & 31 \\
0 & 0 & 1 & 0 & 2 & 0 & 1 & 0 & 3 & 0 & 1 & 0 & 2 & 0 & 1 & 0 & 4 & 0 & 1 & 0 & 2 & 0 & 1 & 0 & 3 & 0 & 1 & 0 & 2 & 0 & 1 & 0 \\
\midrule
-5 & 0 & -5 & -4 & -5 & -3 & -5 & -4 & -5 & -2 & -5 & -4 & -5 & -3 & -5 & -4 & -5 & -1 & -5 & -4 & -5 & -3 & -5 & -4 & -5 & -2 & -5 & -4 & -5 & -3 & -5 & -4 \\
-4 & -4 & 1 & 0 & -4 & -4 & -3 & -3 & -4 & -4 & -2 & -2 & -4 & -4 & -3 & -3 & -4 & -4 & -1 & -1 & -4 & -4 & -3 & -3 & -4 & -4 & -2 & -2 & -4 & -4 & -3 & -3 \\
-3 & -3 & -3 & -3 & 2 & 0 & 1 & 0 & -3 & -3 & -3 & -3 & -2 & -2 & -2 & -2 & -3 & -3 & -3 & -3 & -1 & -1 & -1 & -1 & -3 & -3 & -3 & -3 & -2 & -2 & -2 & -2 \\
-2 & -2 & -2 & -2 & -2 & -2 & -2 & -2 & 3 & 0 & 1 & 0 & 2 & 0 & 1 & 0 & -2 & -2 & -2 & -2 & -2 & -2 & -2 & -2 & -1 & -1 & -1 & -1 & -1 & -1 & -1 & -1 \\
-1 & -1 & -1 & -1 & -1 & -1 & -1 & -1 & -1 & -1 & -1 & -1 & -1 & -1 & -1 & -1 & 4 & 0 & 1 & 0 & 2 & 0 & 1 & 0 & 3 & 0 & 1 & 0 & 2 & 0 & 1 & 0 \\
\midrule
0 & -5 & -4 & -5 & -3 & -5 & -4 & -5 & -2 & -5 & -4 & -5 & -3 & -5 & -4 & -5 & -1 & -5 & -4 & -5 & -3 & -5 & -4 & -5 & -2 & -5 & -4 & -5 & -3 & -5 & -4 & -5 \\
1 & 0 & -4 & -4 & -3 & -3 & -4 & -4 & -2 & -2 & -4 & -4 & -3 & -3 & -4 & -4 & -1 & -1 & -4 & -4 & -3 & -3 & -4 & -4 & -2 & -2 & -4 & -4 & -3 & -3 & -4 & -4 \\
2 & 0 & 1 & 0 & -3 & -3 & -3 & -3 & -2 & -2 & -2 & -2 & -3 & -3 & -3 & -3 & -1 & -1 & -1 & -1 & -3 & -3 & -3 & -3 & -2 & -2 & -2 & -2 & -3 & -3 & -3 & -3 \\
3 & 0 & 1 & 0 & 2 & 0 & 1 & 0 & -2 & -2 & -2 & -2 & -2 & -2 & -2 & -2 & -1 & -1 & -1 & -1 & -1 & -1 & -1 & -1 & -2 & -2 & -2 & -2 & -2 & -2 & -2 & -2 \\
4 & 0 & 1 & 0 & 2 & 0 & 1 & 0 & 3 & 0 & 1 & 0 & 2 & 0 & 1 & 0 & -1 & -1 & -1 & -1 & -1 & -1 & -1 & -1 & -1 & -1 & -1 & -1 & -1 & -1 & -1 & -1 \\
    \bottomrule
  \end{tabular}
  \end{scriptsize}
  \end{center}

\begin{center}
\caption{Receive and send schedules for $p=31$
  ($\skips{i}=1,2,4,8,16,31$, vertical lines) for all processors
  $r,0\leq r<p$.}
\label{tab:p31}
  \begin{scriptsize}
  \begin{tabular}{c|c|cc|cccc|cccccccc|ccccccccccccccc}
    \toprule
0 & 1 & 2 & 3 & 4 & 5 & 6 & 7 & 8 & 9 & 10 & 11 & 12 & 13 & 14 & 15 & 16 & 17 & 18 & 19 & 20 & 21 & 22 & 23 & 24 & 25 & 26 & 27 & 28 & 29 & 30 \\
0 & 0 & 1 & 0 & 2 & 0 & 1 & 0 & 3 & 0 & 1 & 0 & 2 & 0 & 1 & 0 & 4 & 0 & 1 & 0 & 2 & 0 & 1 & 0 & 3 & 0 & 1 & 0 & 2 & 0 & 1 \\
\midrule
-4 & 0 & -5 & -4 & -5 & -3 & -5 & -4 & -5 & -2 & -5 & -4 & -5 & -3 & -5 & -4 & -5 & -1 & -5 & -4 & -5 & -3 & -5 & -4 & -5 & -2 & -5 & -4 & -5 & -3 & -5 \\
-3 & -4 & 1 & 0 & -4 & -4 & -3 & -3 & -4 & -4 & -2 & -2 & -4 & -4 & -3 & -3 & -4 & -4 & -1 & -1 & -4 & -4 & -3 & -3 & -4 & -4 & -2 & -2 & -4 & -4 & -3 \\
-2 & -3 & -3 & -3 & 2 & 0 & 1 & 0 & -3 & -3 & -3 & -3 & -2 & -2 & -2 & -2 & -3 & -3 & -3 & -3 & -1 & -1 & -1 & -1 & -3 & -3 & -3 & -3 & -2 & -2 & -2 \\
-1 & -2 & -2 & -2 & -2 & -2 & -2 & -2 & 3 & 0 & 1 & 0 & 2 & 0 & 1 & 0 & -2 & -2 & -2 & -2 & -2 & -2 & -2 & -2 & -1 & -1 & -1 & -1 & -1 & -1 & -1 \\
-5 & -1 & -1 & -1 & -1 & -1 & -1 & -1 & -1 & -1 & -1 & -1 & -1 & -1 & -1 & -1 & 4 & 0 & 1 & 0 & 2 & 0 & 1 & 0 & 3 & 0 & 1 & 0 & 2 & 0 & 1 \\
\midrule
0 & -5 & -4 & -5 & -3 & -5 & -4 & -5 & -2 & -5 & -4 & -5 & -3 & -5 & -4 & -5 & -1 & -5 & -4 & -5 & -3 & -5 & -4 & -5 & -2 & -5 & -4 & -5 & -3 & -5 & -4 \\
1 & 0 & -4 & -4 & -3 & -3 & -4 & -4 & -2 & -2 & -4 & -4 & -3 & -3 & -4 & -4 & -1 & -1 & -4 & -4 & -3 & -3 & -4 & -4 & -2 & -2 & -4 & -4 & -3 & -3 & -4 \\
2 & 0 & 1 & 0 & -3 & -3 & -3 & -3 & -2 & -2 & -2 & -2 & -3 & -3 & -3 & -3 & -1 & -1 & -1 & -1 & -3 & -3 & -3 & -3 & -2 & -2 & -2 & -2 & -3 & -3 & -3 \\
3 & 0 & 1 & 0 & 2 & 0 & 1 & 0 & -2 & -2 & -2 & -2 & -2 & -2 & -2 & -2 & -1 & -1 & -1 & -1 & -1 & -1 & -1 & -1 & -2 & -2 & -2 & -2 & -2 & -2 & -2 \\
4 & 0 & 1 & 0 & 2 & 0 & 1 & 0 & 3 & 0 & 1 & 0 & 2 & 0 & 1 & -5 & -1 & -1 & -1 & -1 & -1 & -1 & -1 & -1 & -1 & -1 & -1 & -1 & -1 & -1 & -1 \\
    \bottomrule
  \end{tabular}
  \end{scriptsize}
  \end{center}
\end{sidewaystable*}

We have used the schedule computations of
Algorithm~\ref{alg:receiveschedule} and
Algorithm~\ref{alg:sendschedule} to compute the communication
schedules for all $p$ processors for some illustrative values of $p$,
namely $p=20$ in Table~\ref{tab:p20}, $p=33$ in Table~\ref{tab:p33},
$p=32$ in Table~\ref{tab:p32} and $p=31$ in Table~\ref{tab:p31}.
These schedules, in particular the latter three, show the different
schedules around a power of two number of processors, $p=2^5=32$, and
are worth studying in detail: The complexity of the schedule
computations can most likely be improved considerably by exploiting
further structural properties of these schedules. The goal is to be
able to compute both send and receive schedules in $O(\log p)$ time
steps, but whether this is possible is at the moment not known.




\subsection{The broadcast algorithm}
\label{sec:bcastalg}

\begin{algorithm}
  \caption{The $n$-block broadcast algorithm for processor $r,0\leq
    r<p$ of data blocks in array \texttt{buffer}. The count $x$ is the
    number of empty first rounds.  Blocks smaller than $0$ are neither
    sent nor received, and for blocks larger than $n-1$, block $n-1$
    is sent instead.}
  \label{alg:broadcast}
  \begin{algorithmic}[1]
    \State$\mathtt{recvsched}\gets\Call{recvsched}{r,q}$
    \State$\mathtt{sendsched}\gets\Call{sendsched}{r}$
    \\
    \State $x\gets (q-(n-1+q)\bmod q)\bmod q$ \Comment Dummy rounds
    \For{$i=0,1,\ldots,q-1$}
    \If{$i<x$} \Comment Adjust schedule, $x$ virtual rounds already done
    \State$\mathtt{recvsched}[i]\gets\mathtt{recvsched}[i]-x+q$
    \State$\mathtt{sendsched}[i]\gets\mathtt{sendsched}[i]-x+q$
    \Else
    \State$\mathtt{recvsched}[i]\gets\mathtt{recvsched}[i]-x$
    \State$\mathtt{sendsched}[i]\gets\mathtt{sendsched}[i]-x$
    \EndIf
    \EndFor    
    \State $i\gets x$
    \While{$i<n+q-1+x$}
    \State $k\gets i\bmod q$
    \State $t^k\gets (r+\skips{k})\bmod p$ \Comment to- and from-processors
    \State $f^k\gets (r-\skips{k}+p)\bmod p$
    \\
    \State
    $\bidirec{\mathtt{buffer}[\mathtt{sendsched}[k]],t^k}{\mathtt{buffer}[\mathtt{recvsched}[k]],f^k}$
    \\
    \State $\mathtt{sendsched}[k]\gets \mathtt{sendsched}[k]+q$
    \State $\mathtt{recvsched}[k]\gets \mathtt{recvsched}[k]+q$
    \\
    \State $i\gets i+1$
    \EndWhile
\end{algorithmic}
\end{algorithm}

Finally, the \texttt{sendsched} and \texttt{recvsched} schedules can
be used to broadcast the $n$ data blocks from root processor $r=0$ to
all other processor in the promised $n-1+q$ communication rounds.  The
full broadcast algorithm is shown as Algorithm~\ref{alg:broadcast},
and can readily be used to implement the \mpibcast operation of
MPI~\cite[Chapter 5]{MPI-3.1}. Each processor first computes its send
and receive schedules, and then goes through the $n-1+q$ communication
rounds. The rounds are divided into phases of $q$ rounds. In each
round $k=i\bmod q$ in a phase, in parallel (indicated by $\parallel$)
a processor receives and sends a data block from/to from- and
to-processors ($t^k$ and $f^k$) according to its receive and send
schedules. For the next phase, the block schedules are incremented by
$q$. Data are assumed to be stored block wise in the array
\texttt{buffer}.

The schedules assume a number of rounds that is a multiple of $q$. In
order to achieve that, we introduce a number of empty dummy rounds $x$
such that $x+n-1+q$ is a multiple of $q$ and send and receive
(virtual) negative blocks $-x,-x+1,\ldots,-1$ for the $x$ first
rounds. Send and receive blocks from the schedules that are negative
are neither sent nor received, and blocks that are larger than $n-1$
are treated as block $n-1$. Therefore, actually nothing is done in the
$x$ dummy rounds. This trick is from~\cite{Traff08:optibcast}.

Assuming a homogeneous, linear-cost communication model in which
communication of $m$ units of data between any two neighboring
processors in the circulant graph takes $\alpha+\beta m$ time units,
the number of blocks $n$ in which to divide $m$ and the ensuing block
size $m/n$ can be balanced against each other to achieve a best
possible (under this model) broadcasting time using the round-optimal
schedule. We summarize the result in the following theorem.

\begin{theorem}
  In a fully connected, $1$-ported communication network with $p$
  processors with homogeneous, linear communication costs with latency
  $\alpha$ and time per unit $\beta$, broadcasting from a root
  processor to all other processors of data of size $m$ can be done in
  time $\alpha\ceiling{\log_2 p-1} + 2\sqrt{\ceiling{\log_2
      p-1}\alpha\beta m}+ \beta m$ with a schedule construction
  overhead of $O(\log^3 p)$.
\end{theorem}

\subsection{A regular allgather algorithm}
\label{sec:allgatheralg}

Before applying the broadcast schedules to the irregular
all-to-all-broadcast (allgather) problem, we observe that the circulant
graph communication structure can easily be used to solve the regular
allgather problem, in which each processor has a data block of the
same size $m/p$. This follows from the observations of
Lemma~\ref{lem:sizes}. The algorithm goes through $q$ communication
rounds, using the pattern of to- and from-processors as in the
broadcast algorithm in Algorithm~\ref{alg:broadcast}. In each round
$k=0,\ldots q-1$, processor $r$ sends a range of
$\skips{k+1}-\skips{k}$ blocks to the to-processor and receives an
adjacent range of $\skips{k+1}-\skips{k}$ blocks from its
from-processor. By this, the algorithm maintains the invariant
that after round $k$, each processor will have a successive range of
$\skips{k}+(\skips{k+1}-\skips{k})=\skips{k+1}$ blocks.
After the $q$ rounds, each processor will have
its own block and $p-1$ blocks from the other processors. The
algorithm is shown as Algorithm~\ref{alg:allgather}. Note that each
processor stores its blocks in a \texttt{buffer} array indexed from
$0$, with $\mathtt{buffer}[0]$ storing initially the block contributed
by the processor (the ordering convention of MPI is different). It is
conveneint the use the edges of the circulant graph communication
structure in the opposite direction of the broadcast algorithm in
Algorithm~\ref{alg:broadcast}.

\begin{algorithm}
  \caption{The regular allgather for processor $r,0\leq r<p$. The data
    blocks of size $m/p$ are stored consecutively in array an
    \texttt{buffer} and $\mathtt{buffer}[0]$ is the inital block of
    size $m/p$ for processor $r$. Note that the edges of the circulant
    graph communication structure are used in the opposite direction of
    the broadcast algorithm.}
  \label{alg:allgather}
  \begin{algorithmic}[1]
    \For{$k=0,1,\ldots,q-1$}
    \State $t^k\gets (r-\skips{k}+p)\bmod p$ \Comment to- and from-processors
    \State $f^k\gets (r+\skips{k})\bmod p$ 
    \\
    \State
    $\bidirec{\mathtt{buffer}[0,\ldots,\skips{k+1}-\skips{k}-1],t^k}{\mathtt{buffer}[\skips{k},\ldots,\skips{k+1}-1],f^k}$
    \EndFor
\end{algorithmic}
\end{algorithm}

Assuming linear time communication costs, the complexity of the
algorithms is $\alpha\ceiling{\log_2 p}+\beta (p-1)m/p$ since the
number of communicated blocks of size $m/p$ roughly doubles per round
and sums to $p-1$. This observation is an alternative to the
well-known dissemination allgather algorithm in~\cite{Bruck97}. 

\begin{algorithm}
  \caption{The census algorithm for processor $r,0\leq r<p$ with an
    associative, commutative operator $\oplus$. The input for each
    processor $r$ is stored in $x_r$, and the goal of the algorithm is
    to compute for each processor $r$ the sum
    $S=\oplus_{r=0}^{p-1}x_r$. For convenience, a neutral element
    $\bot$ for $\oplus$ is assumed; this is not needed for a concrete
    implementation. Note that the edges of the circulant 
    graph communication structure are used in the opposite direction of the
    broadcast algorithm; edges get adjusted at each round of the
    algorithm.}
  \label{alg:allreduce}
  \begin{algorithmic}[1]
    \State $S\gets\bot$
    \For{$k=0,1,\ldots,q-1$}
    \Comment to- and from-processors
    \If{$\skips{i}+\skips{i}>\skips{i+1}$}
    \State $t^k\gets (r-\skips{k}+1+p)\bmod p$ 
    \State $f^k\gets (r+\skips{k}-1)\bmod p$
    \State $O\gets S$
    \Else
    \State $t^k\gets (r-\skips{k}+p)\bmod p$ 
    \State $f^k\gets (r+\skips{k})\bmod p$  
    \State $O\gets x_r\oplus S$
    \EndIf
    \State $\bidirec{O,t^k}{B,f^k}$
    \State $S\gets S\oplus B$
    \EndFor
    \State $S\gets x_r\oplus S$
  \end{algorithmic}
\end{algorithm}

The same observations can be applied to the census (allreduce) problem
and give rise to much the same algorithm as
in~\cite{BarNoyKipnisSchieber93}. Here, an associative and commutative
operator $\oplus$ is given, and each processor $r,0\leq r<p$ has an
input value $x_r$. The problem is to compute, in some order, for each
processor the sum $\oplus_{r=0}^{p-1}x_r$ of all the input values.
The idea is to maintain as invariant for each processor $r,0\leq r<p$
after communication round $k$ the sum
$S=\oplus_{i=1}^{\skips{k+1}-1}x_{(r+i)\bmod p}$ of the values of the
$\skips{k+1}-1$ next (higher ranked, modulo the number processors)
processors, excluding the value $x_r$ for processor $r$ itself. The
invariant can be established initially by setting $S$ to the neutral
element $\bot$ for $\oplus$ (a neutral element is strictly not needed,
the implementation just need to be a bit more careful for the first
communication round). To establish the invariant for round $k+1$ there
are two possibilities. If $\skips{k}+\skips{k}=\skips{k+1}$ (see
Lemma~\ref{lem:sizes}), the partial sum $S$ for processor $r$ and the
partial sum from from-processor $r'=(r+\skips{k})\bmod p$ plus the
value $x_{r'}$ needs to be summed, so processor $r$ receives
$x_{r'}\oplus S$ from its from-processor $r'$. If
$\skips{k}+\skips{k}=\skips{k+1}+1>\skips{k+1}$ (see
Lemma~\ref{lem:sizes}, note that these conditions are equivalent to
checking whether $\skips{i+1}$ is odd or even), processor $r'$ is of
no help, and the processor instead receives from the processor just
before from-processor $r'$ its value of $S$, and adds that to its own
partial sum. By the invariant, after the last iteration,
$S=\oplus_{i=1}^{\skips{q}-1}x_{(r+i)\bmod
  p}=\oplus_{i=1}^{p-1}x_{(r+i)\bmod p}$, so to get the final result,
$S$ has to be added to $x_r$.  The algorithm is shown as
Algorithm~\ref{alg:allreduce}.

Assuming linear time communication costs, the complexity of the
algorithm is $\ceiling{\log_2 p}(\alpha+\beta m)$ where $m$ is the
size of the input elements $x_r$, since in each of the $q$ rounds a
partial sum of size $m$ is sent and received by each processor. The
algorithm can be useful for cases where $\beta m$ is small compared to
$\alpha$.

\subsection{The irregular allgather algorithm}
\label{sec:allgathervalg}

\begin{algorithm}
  \caption{The $n$-block irregular allgather algorithm for processor
    $,r,0\leq r<p$ for data in the arrays $\mathtt{buffer}[j],0\leq
    j<p$. The count $x$ is the number of empty first rounds.  Blocks
    smaller than $0$ are neither sent nor received, and for blocks
    larger than $n-1$, block $n-1$ is sent instead.}
  \label{alg:irregallgather}
  \begin{algorithmic}[1]
    \For{$j=0,1,\ldots,p-1$}
    \State$\mathtt{recvsched}[j]\gets\Call{recvsched}{j,q}$
    \EndFor
    \For{$j=0,1,\ldots,p-1$}
    \For{$k=0,1,\ldots,q-1$}
    \State $t^k\gets (j+\skips{k})\bmod p$
    \State$\mathtt{sendsched}[j][k]\gets\mathtt{recvsched}[t^k][k]$
    \EndFor
    \EndFor    
    \\
    \State $x\gets (q-(n-1+q)\bmod q)\bmod q$ \Comment Dummy rounds
    \State $i\gets x$
    \While{$i<n+q-1+x$}
    \State $k\gets i\bmod q$
    \State $t^k\gets (r+\skips{k})\bmod p$ \Comment to- and from-processors
    \State $f^k\gets (r-\skips{k}+p)\bmod p$
    \\
    \For{$j=0,1,\ldots,p-1$} \Comment Pack
    \State $r'\gets (r-j+p)\bmod p$
    \State $\mathtt{tempin}[j]\gets\mathtt{buffer}[j][\mathtt{sendsched}[r'][k]]$
    \State $\mathtt{sendsched}[r'][k]\gets \mathtt{sendsched}[r'][k]+q$
    \EndFor

   \State $\bidirec{\mathtt{tempin},t^k}{\mathtt{tempout},f^k}$

    \For{$j=0,1,\ldots,p-1$} \Comment Unpack
    \State $r'\gets (r-j+p)\bmod p$
    \State $\mathtt{buffer}[j][\mathtt{recvsched}[r'][k]]\gets\mathtt{tempout}[j]$
    \State $\mathtt{recvsched}[r'][k]\gets \mathtt{recvsched}[r'][k]+q$
    \EndFor
    
    \State $i\gets i+1$
    \EndWhile
\end{algorithmic}
\end{algorithm}

As a new application of the symmetric, circulant graph communication
pattern and the broadcast send and receive schedules, we give an
algorithm for the symmetric all-to-all-broadcast (allgather) operation in
which all processors have a buffer of data to be broadcast to all
other processors, in particular for the irregular case where the
buffers may be of different sizes for different processors. This
collective operation is likewise found in the MPI standard as
\mpiallgatherv~\cite[Chapter 5]{MPI-3.1}. The algorithm is shown as
Algorithm~\ref{alg:irregallgather}.  Extending the broadcast algorithm in
Algorithm~\ref{alg:broadcast}, data are stored in an array of $p$
buffers, one for each processor, $\mathtt{buffer}[j],0\leq
j<p$. Processor $r$ initially has data in $\mathtt{buffer}[r]$ which
it shall broadcast to the other processors, and will receive data from
all other processors in $\mathtt{buffer}[j],j\neq r$. The buffers may
be of different sizes, which we do not (have to) keep track of
explicitly.

The idea is simple. Simply let each processor play the role of a root
processor, and compute the $p$ receive schedules accordingly. The way
this idea is actually realized is slightly different. First we compute
a \emph{full schedule} consisting of the receive schedules for all
processors $j,0\leq j<p$. The schedule $\mathtt{recvsched}[j]$ thus
gives the blocks that a virtual processor $i$ has to receive when a
virtual processor $0$ is the broadcast root. A processor $r, 0\leq
r<p$ therefore receives into $\mathtt{buffer}[j],0\leq j<p$ the blocks
determined by $\mathtt{recvsched}[(r-j+p)\bmod p]$. The blocks that
are to be sent are determined likewise.

Unfortunately, the blocks that are sent and received in each round
come from different arrays $\mathtt{buffer}[j],0\leq j<p$ and are
therefore not consecutive in memory. Instead of assuming that send and
receive communication operations can deal immediately with
non-consecutive data, we explicitly pack and unpack the blocks for
each round to and from consecutive temporary buffers of blocks. A
practical implementation of the algorithm will have to deal with this
potential overhead, that can possibly be at least partly overlapped
with the actual communication and/or otherwise minimized.

The total amount of data $m=\sum_{r=0}^{p-1}m_r$ to be
all-to-all-broadcast is the sum of the sizes of data $m_r$ initially
contributed by each processor $r$. From this, the number of
communication rounds $n$ is then computed, so as to minimize the
overall communication time (see Section~\ref{sec:results}). For each
$\mathtt{buffer}[j]$ the size per block is the buffer size divided by
$n$.

Since only receive schedules need to be computed explicitly, the total
time for the full schedule precomputation is $O(p\log^2 p)$; therefore
the algorithm can be effective only for $m$ beyond a certain threshold
that can amortize this overhead. This overhead can at least
theoretically be improved to $O(\log^2 p+p\log p)$ by letting each
processor compute only the schedule for itself (as in the broadcast
algorithm), and then allgathering these schedules on all processors.
(assuming a $\log p$ round allgather algorithm with linear
communication volume per processor~\cite{Bruck97} or as just explained in
Section~\ref{sec:allgatheralg}).

We summarize the discussion in the following, new theorem.

\begin{theorem}
  In a fully connected, $1$-ported communication network with $p$
  processors with homogeneous, linear communication costs with latency
  $\alpha$ and time per unit $\beta$, the irregular all-to-all-broadcast
  (allgather) problem in which each processor contributes data of size
  $m_r$ for a total problem size of $m=\sum_{r=0}^{p-1}m_r$
  can be solved in time $\alpha\ceiling{\log_2 p-1} +
  2\sqrt{\ceiling{\log_2 p-1}\alpha\beta m}+ \beta m$ with a schedule
  construction overhead of $O(p\log^2 p)$.
\end{theorem}

\section{Implementation and Experimental Results}
\label{sec:results}


The schedule computations and broadcast and irregular allgather
algorithms have been implemented in C for supporting the MPI
collectives \mpibcast and \mpiallgatherv~\cite[Chapter 5]{MPI-3.1}
(code is available from the author, and the schedule computations has
been exhaustively verified up to a large $p>100\, 000$).  The
The implementations were tested and benchmarked on a small cluster
with 36~dual socket compute nodes, each with an Intel(R) Xeon(R) Gold
6130F 16-core processor. The nodes are interconnected via dual Intel
Omnipath interconnects each with a bandwidth of \SI{100}{\Gbps{}}. The
implementations and benchmarks were compiled with \gcc with the
\texttt{-O3} option.  For \mpibcast, we give results for three
different MPI libraries, namely \hydrampich, \hydraopenmpi, and
\hydraintelmpi. For \mpiallgatherv we show results with the presumably
best performing MPI library at disposal, \hydraintelmpi.  We report
the performance of our implementations against the native \mpibcast
and \mpiallgatherv operations of each of these libraries.  The
benchmark measures time for the operations individually for increasing
problem sizes, for each measurement records the time for the slowest
MPI process, and reports the minimum time over a series of
\repetitions~measurements~\cite{HunoldCarpenamarie16}. Our comparison
is fair, in the sense that the times reported for our implementations
include both schedule computation and communication time. Our
implementations have the same interface and functionality as the
native \mpibcast and \mpiallgatherv operations, but do not cater for
technicalities related to correct handling of MPI user-defined
datatypes (non-contiguous data buffers).

Results are reported for $p=N\times n$ MPI processes where $N$ is the
number of nodes and $n$ the number of processes per node. The input of
size $m$ is in a consecutive buffer of $m/\texttt{sizeof(int)}$ MPI
integers of MPI type \texttt{MPI\_INT}.  Here it is important to
stress that our implementations are done under the assumption of
homogeneous communication costs (same between any pair of processes),
which is not realistic for a clustered system where the cost of
communication inside compute nodes is typically different (in certain
respects, lower) than the cost of communication between processes on
different compute nodes. There are several ways of adapting broadcast
and allgather implementations to non-homogeneous, hierarchical
settings, see for instance~\cite{Traff20:mpidecomp}, but this is
future work not considered here.


\subsection{Broadcast results}

For the broadcast implementation, a ``best possible'' block size for
dividing the input of size $m$ (Bytes) into $n$ blocks is estimated as
$F\sqrt{m/\ceiling{\log p}}$ for a constant factor $F$ that has been chosen
experimentally. We make no claim that the choice for $F$ is indeed
best possible, and also not that a best possible block size can be
determined by such a simple expression. Different choices of $F$ were
selected for different MPI libraries. We have performed measurements
with $m$ ranging from 1 to $100\,000\,000$ MPI integers ($400\,000\,000$
Bytes) by multiplying alternatingly by $2$ and $5$.

\begin{figure*}
  \begin{center}
    \includegraphics[width=.3\linewidth]{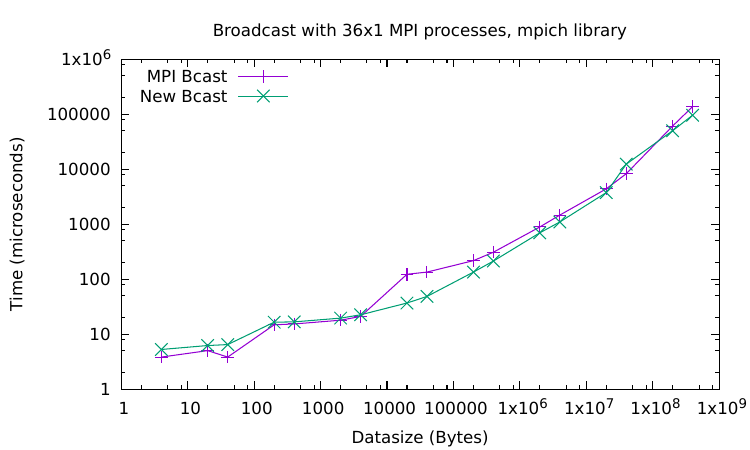}
    \includegraphics[width=.3\linewidth]{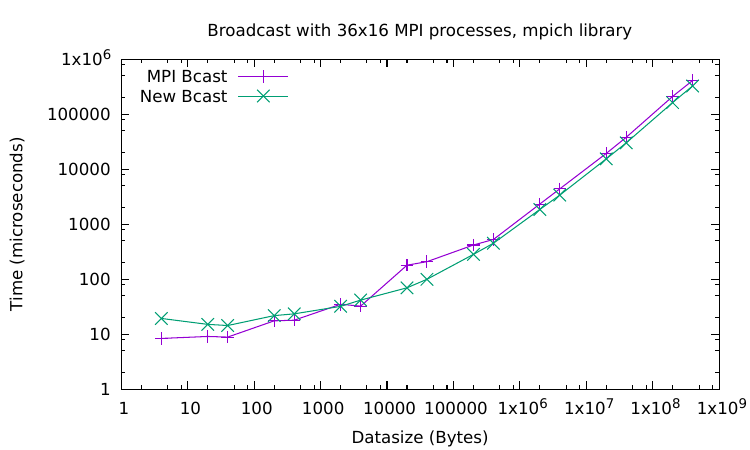}
    \includegraphics[width=.3\linewidth]{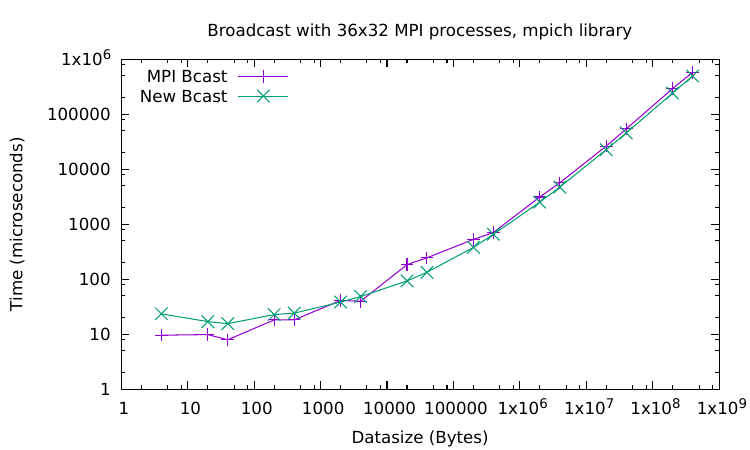}
  \end{center}
  \caption{Broadcast results, native versus new, with the \hydrampich
    library with $36\times 1$, $36\times 16$ and $36\times 32$ MPI
    processes.  The constant factor $F$ for the block size has been chosen as
    $F=80$.}
  \label{fig:bcastmpich}
\end{figure*}

\begin{figure*}
  \begin{center}
    \includegraphics[width=.3\linewidth]{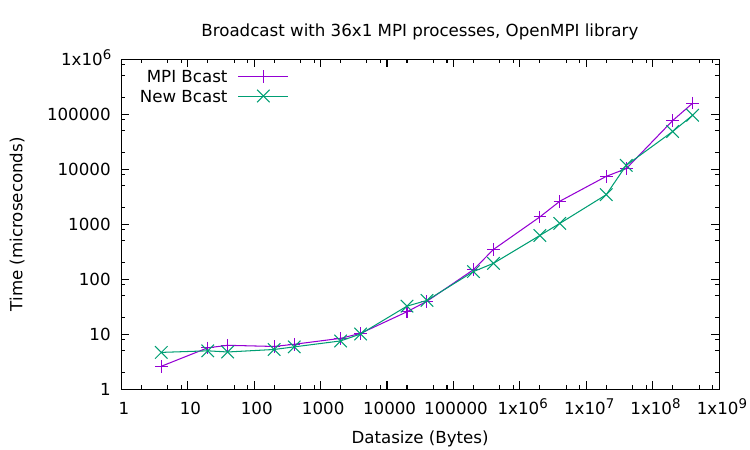}
    \includegraphics[width=.3\linewidth]{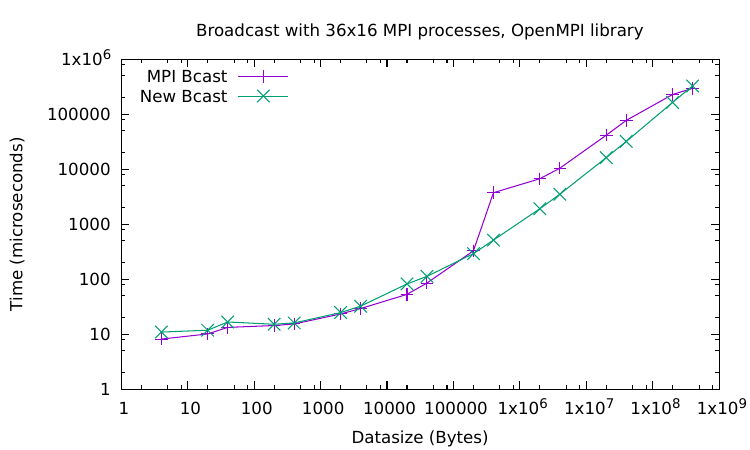}
    \includegraphics[width=.3\linewidth]{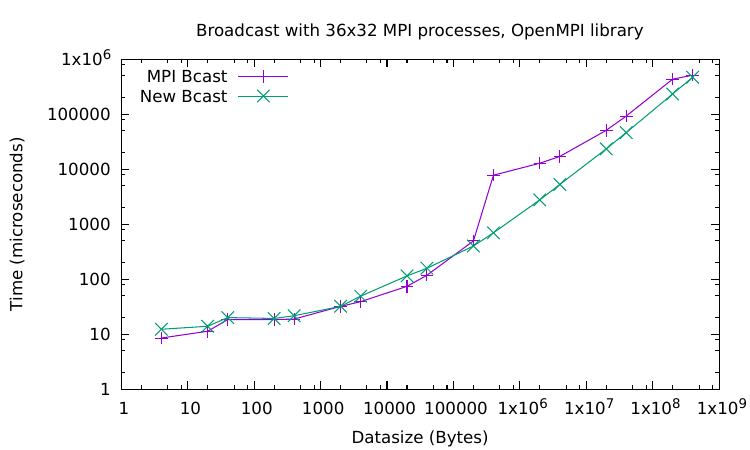}
  \end{center}
  \caption{Broadcast results, native versus new, with the \hydraopenmpi
    library with $36\times 1$, $36\times 16$ and $36\times 32$ MPI
    processes. The constant factor $F$ for the block size has been chosen as
    $F=100$.}
  \label{fig:bcastopenmpi}
\end{figure*}

\begin{figure*}
  \begin{center}
    \includegraphics[width=.3\linewidth]{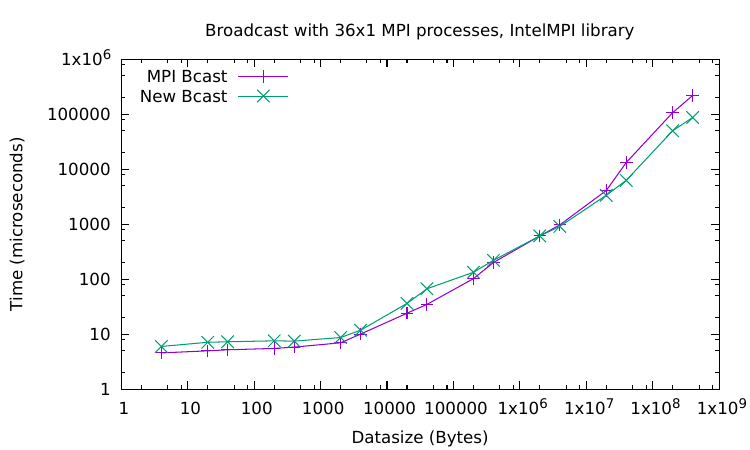}
    \includegraphics[width=.3\linewidth]{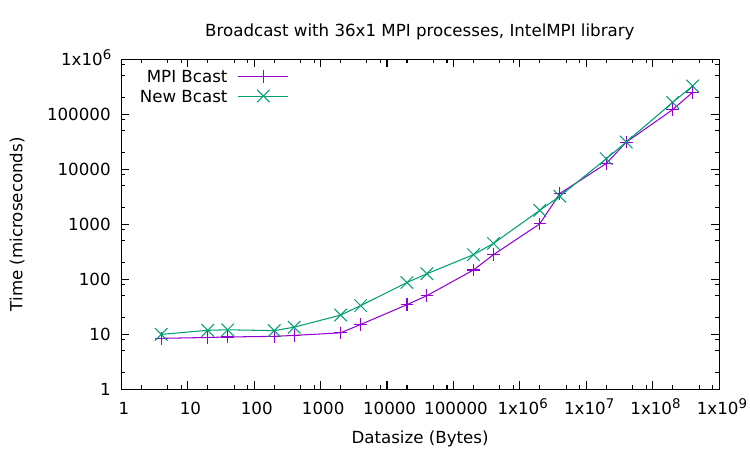}
    \includegraphics[width=.3\linewidth]{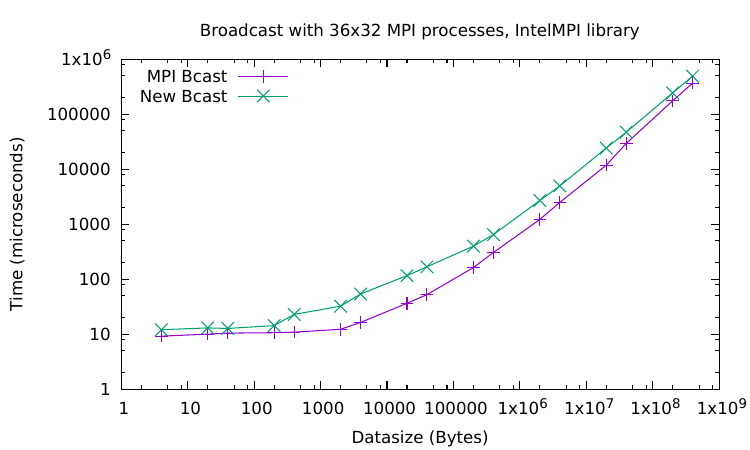}
  \end{center}
  \caption{Broadcast results, native versus new, with the
    \hydraintelmpi library with $36\times 1$, $36\times 16$ and
    $36\times 32$ MPI processes. The constant factor $F$ for the block
    size has been chosen as $F=120$.}
  \label{fig:bcastintelmpi}
\end{figure*}

The broadcast results with the \hydrampich library can be seen in
Figure~\ref{fig:bcastmpich}, with the \hydraopenmpi library in
Figure~\ref{fig:bcastopenmpi}, and with the \hydraintelmpi library in
Figure~\ref{fig:bcastintelmpi}. The plots are doubly logarithmic, so
significant improvements and differences in performance may appear
smaller than they actually are. The performance of the library native
\mpibcast operation is plotted against the implementation using the
optimal schedule computation (New Bcast).  We have run with the
maximum number of nodes of the cluster $N=36$, and number of MPI
processes per node $n=1,16,32$. As can be seen, the library native
\mpibcast operation performs quite differently for different
libraries. For $m=400\,000\,000$ Bytes and $n=1$, the \hydrampich
library is the fastest with about \SI{140}{\ms} and \hydraintelmpi the
slowest with \SI{220}{\ms}. The new broadcast implementation
is in all cases faster with \SI{95}{\ms} and \SI{87}{\ms},
respectively. When the nodes are full, $n=32$, the picture
changes. Here the \hydraintelmpi library is the fastest with
\SI{368}{\ms}, and the new implementation actually slower
with about \SI{495}{\ms}. For the other libraries, the new
implementation is faster than the native \mpibcast by a small margin
(\SI{470}{\ms} against \SI{512}{\ms} for the \hydraopenmpi
library). The \hydraintelmpi library seems to have a better
hierarchical implementation for clustered systems. For smaller problem
sizes, the new implementation can in many data size ranges significantly
outperform the native \mpibcast, sometimes by a factor of two to
three.

The results indicate considerable potential for using the
round-optimal broadcast algorithm in MPI libraries (as
in~\cite{Traff06:mpisxcoll}) as the basis for supporting \mpibcast,
but engineering is needed for determining good block sizes and
reducing overheads.

\subsection{Irregular allgather results}
\label{sec:allgatherv}

The experimental results on the irregular allgather implementation are
cursory and not exhaustive. We create a simple instance of an
irregular problem, where different MPI processes have blocks of
different sizes to broadcast to the other MPI processes. We let $m_r$
be the size of the data to be broadcast by process $r$, except for the
last process $r=p-1$, where we take $m_{p-1}=m-\sum_{r=0}^{p-2}m_r$
for some given total problem size $m$. This way, the problem size can
be made exactly comparable to the problem sizes used in the broadcast
experiments, for instance.  For the simple experiment here, we have
chosen $m_r=(r\bmod 3)\floor{m/p}, 0\leq r<p-1$ such that for a chosen $m$,
$m=\sum_{r=0}^{p-1}m_r$.

For the allgather implementation, a ``best possible'' number of blocks
$n$ into which the input of size $m$ (Bytes) is divided is estimated
as $\sqrt{m\ceiling{\log p}}/G$ for another constant factor $G$ that has been
chosen experimentally. Also here, we make no claim that the choice for
$G$ is indeed best possible, and also not that a best possible number
of blocks can be determined by such a simple expression. We 
report experiments only with the \hydraintelmpi library which show
both the promise and the problems with the current implementation.  As
for broadcast, we have performed measurements with $m$ ranging from 1
to $100\,000\,000$ MPI integers ($400\,000\,000$ Bytes) by multiplying
alternatingly by $2$ and $5$.

\begin{figure*}
  \begin{center}
    \includegraphics[width=.3\linewidth]{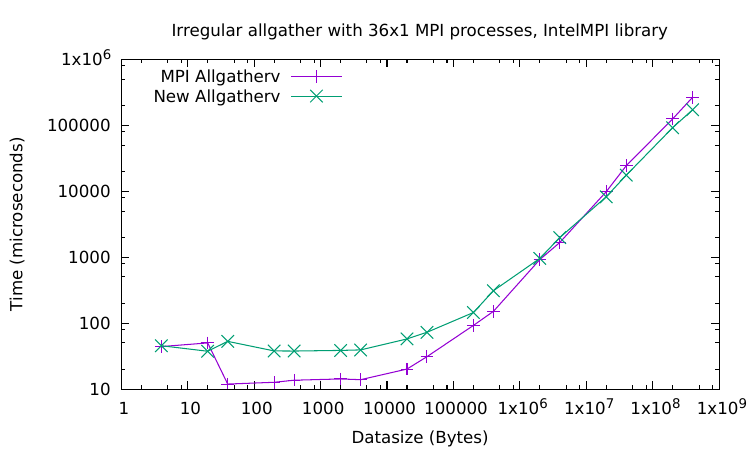}
    \includegraphics[width=.3\linewidth]{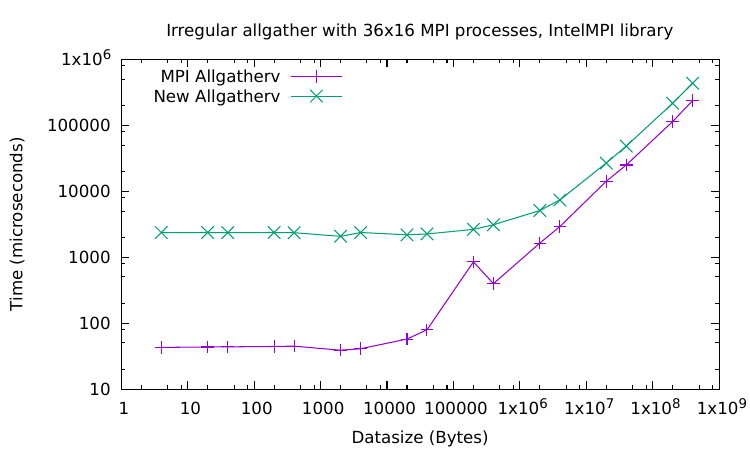}
    \includegraphics[width=.3\linewidth]{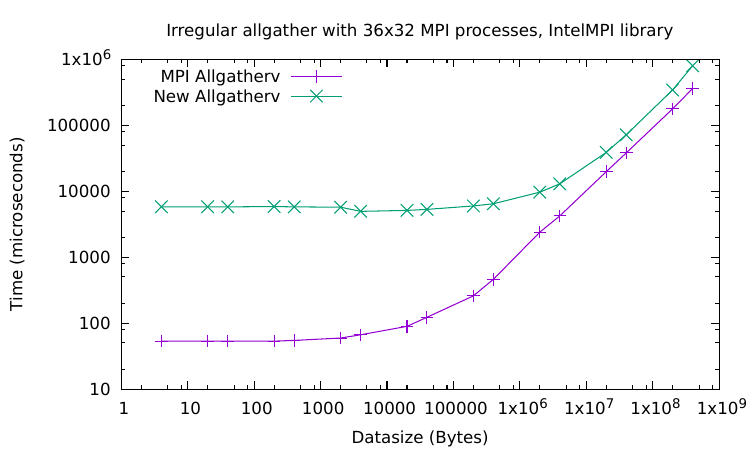}
  \end{center}
  \caption{Irregular allgather results, native versus new, with the
    \hydraintelmpi library with $36\times 1$, $36\times 16$ and $36\times 32$
    MPI processes. The constant factor $G$ for the number of blocks has been
    chosen as $G=40$.}
  \label{fig:allgatintelmpi}
\end{figure*}

The irregular allgather results with the \hydraintelmpi library are
found in Figure~\ref{fig:allgatintelmpi}. With only one MPI process
per compute node, $n=1$, the new algorithm outperforms the native
\mpiallgatherv operation with about \SI{171}{\ms} against
\SI{265}{\ms}. This is a promising result (which of course needs to be
corrobated with experiments with diverse, irregular problems and
distributions). As expected, this unfortunately changes quite
dramatically with more than one MPI process per compute node where the
new implementations is about twice as slow as the native
\mpiallgatherv when $m$ is large. The overhead in this case likely
mostly stems from the packing and unpacking, which makes it necessary
to go over the total data twice (in addition to the communication).
For small problem sizes, when the number of processes increase from 36
to 576 to 1152, the schedule computation overhead becomes considerable
and dominant, from around \SI{40}{\microsecond} to about
\SI{5800}{\microsecond}. This is consistent with the $O(p\log^3 p)$
complexity of the schedule computation and the 32-fold increase in $p$.
The schedule computation overhead could be amortized by precomputing
the schedule (at high space cost), or part of the schedule as
suggested in Section~\ref{sec:allgatherv}. The packing and unpacking
problem in the implementation (and algorithm as explained in
Section~\ref{sec:allgatherv}) might be alleviated some, possibly by
the use of MPI derived datatypes, but more engineering work is needed
here. More experimental work is called for with different types of
irregular problems. It would also be worthwhile to compare against the
irregular allgather algorithm described in~\cite{Traff10:largeallgat}.

\section{Open Problems}

\begin{table*}
\begin{center}
  \caption{Two different receive and send schedules for $p=9$ processors.
    The differences are highlighted in the right schedule.}
  \label{tab:p9}
  \begin{small}
  \begin{tabular}{c|c|c|cc|cccc}
    \toprule
0 & 1 & 2 & 3 & 4 & 5 & 6 & 7 & 8 \\
0 & 0 & 1 & 2 & 0 & 3 & 0 & 1 & 2 \\
\midrule
-2 & 0 & -4 & -3 & -2 & -4 & -1 & -4 & -3 \\
-3 & -2 & 1 & -4 & -3 & -2 & -2 & -1 & -4 \\
-1 & -3 & -2 & 2 & 0 & -3 & -3 & -2 & -1 \\
-4 & -1 & -1 & -1 & -1 & 3 & 0 & 1 & 2 \\
\midrule
0 & -4 & -3 & -2 & -4 & -1 & -4 & -3 & -2 \\
1 & -4 & -3 & -2 & -2 & -1 & -4 & -3 & -2 \\
2 & 0 & -3 & -3 & -2 & -1 & -1 & -3 & -2 \\
3 & 0 & 1 & 2 & -4 & -1 & -1 & -1 & -1 \\
\bottomrule
  \end{tabular}
  \hspace{1cm}
  \begin{tabular}{c|c|c|cc|cccc}
    \toprule
0 & 1 & 2 & 3 & 4 & 5 & 6 & 7 & 8 \\
0 & 0 & 1 & 2 & 0 & 3 & 0 & 1 & 2 \\
\midrule
-2 & 0 & -4 & -3 & -2 & -4 & -1 & -4 & -3 \\
-3 & -2 & 1 & -4 & -3 & -2 & -2 & -1 & \textbf{-1} \\
-1 & -3 & -2 & 2 & 0 & -3 & -3 & -2 & \textbf{-4} \\
-4 & -1 & -1 & -1 & -1 & 3 & 0 & 1 & 2 \\
\midrule
0 & -4 & -3 & -2 & -4 & -1 & -4 & -3 & -2 \\
1 & -4 & -3 & -2 & -2 & -1 & \textbf{-1} & -3 & -2 \\
2 & 0 & -3 & -3 & -2 & \textbf{-4} & -1 & -3 & -2 \\
3 & 0 & 1 & 2 & -4 & -1 & -1 & -1 & -1 \\
\bottomrule
  \end{tabular}
  \end{small}
  \end{center}
\end{table*}

The construction given for finding a correct receive schedule is in a
sense quite greedy, and despite being sublinear in $p$, relatively
expensive. The major open question is how much this can be improved,
especially whether it is possible to compute both send and receive
schedules in $O(\log p)$ time steps per processor? The author believes
so. Another interesting questions is whether the schedules are unique
(up to simple permutations) for the given, circulant graph
structure. Here the answer is actually ``no'' as can be seen by the
counterexample for $p=9$ processes in shown in Table~\ref{tab:p9}. How
many schedules are there then for given $p$?  This looks like a
non-trivial combinatorial problem, which it might be enlightening to
solve.

\section{Summary}

This paper showed that round-optimal send and receive schedules for
broadcasting with a fully symmetric circulant graph communication
pattern in fully-connected, bidirectional, $1$-ported, $p$-processor
communication networks can indeed be computed efficiently in
sublinear, $O(\log^3 p)$ time steps for each of the $p$ processors,
and without any communication between the processors. This is highly
desirable for using such schedules for practically efficient broadcast
implementations in communication libraries like
MPI~\cite{MPI-3.1}. The schedule computations have been implemented
and indeed used for broadcast and irregular allgather with
considerable improvements over native MPI library implementations for
the corresponding \mpibcast and \mpiallgatherv operations.  The
interesting, open theoretical challenge (with practical impact)
remains to do the schedule computation even more efficiently, possibly
in only $O(\log p)$ steps.

\bibliographystyle{plain}
\bibliography{traff,parallel} 

\end{document}